# Causality as a unifying approach between activation and connectivity analysis of fMRI data

Nevio Dubbini

Laboratory of Clinical Biochemistry and Molecular Biology, University of Pisa, Italy
*Correspondence: Nevio Dubbini, Laboratory of Clinical Biochemistry and Molecular Biology, University of Pisa, Via Roma 67 - 56126, Pisa, Italy; e-mail: nevio.dubbini@for.unipi.it

**Abstract:** This paper indicates causality as the tool that unifies the analysis of both activations and connectivity of brain areas, obtained with fMRI data. Causality analysis is commonly applied to study connectivity, so this work focuses on demonstrating that also the detection of activations can be handled with a causality analysis. We test our method on finger tapping data, in which GLM and Granger Causality approaches are compared in finding activations. Granger causality not only performs the task well, but indeed we obtained a better localization (i.e. precision) of activations. As a result we claim that causality must be the main tool to investigate activations, since it is a measure of ``how much'' the stimulus influences the BOLD signal, and since it unifies connectivity and activations analysis under the same area.
*keywords*: Causality, activations, functional connectivity, fMRI

# 1 Introduction

The application of MRI to the imaging of the human brain in a completely non invasive way (functional magnetic resonance imaging - fMRI) is a rapidly expanding field of research. fMRI exploits blood oxygenation level dependent (BOLD) contrast, which utilizes the different magnetic properties of oxygenated (diamagnetic) and deoxygenated (paramagnetic) blood. These differences result in detectable changes in image intensity, but relatively low signal-to-noise ratio, head movements, or involuntary muscles activity make detection of the activation-related signal changes difficult ([Nol]). On the other hand, available methodology allows the acquisition of data sets with hundreds of images per voxel so that data can be statistically analyzed.

The analysis of fMRI data is not straightforward, because of the presence of noise combined with the small size of signal changes, but also because the BOLD signal of a voxel is the result of the activity of thousands of neuronal cells. Another controversial related issue is the relation between fMRI BOLD signal change and neuronal activity ([Heg]). It is clear that fMRI signals arise from changes in local hemodynamics, that in turn result from alterations in neuronal activities, but how neuronal activities, hemodynamic responses, and BOLD signal changes are related is not fully defined. Many unknowns affect the knowledge of this relation. fMRI signal might reflect changes in neuronal synchrony without a concomitant increase in mean firing rate [Fri]. Similarly, the modulation in fMRI responses can be caused by either large changes in the firing rates in a small sub population of neurons (inside a voxel), or small changes in the firing rates in a much larger subpopulation of neurons [Sca]. In [Rai] it is asserted that it is not correct to use functional neuroimaging signals as simple surrogate measures of the spiking activities of neurons, but it would be better to look at these signals as complementary information related to the input into a neuronal assembly, which may [Muk] or may not [Lau,Tho] correlate with the output.

According to these facts, statistical and model based techniques to analyze fMRI data are not only desirable, but also necessary. In the vast majority of brain studies with fMRI the two main purposes of data analysis are the identification of areas 'activated' in response to a task, and the detection of the connectivity relationships among these areas. The most used method to detect neural activations is the General Linear Model (GLM), in which the parameters of a linear model are estimated with a regression-based technique. Connectivity is mainly investigated by using causality measures. Granger Causality (GC) is one among the most applied methods [Bli,Gra].

In this paper, we propose the use of causality measures to detect activations of brain areas, as revealed from fMRI data. The aim is to achieve a unifying approach between activation and connectivity analysis under the area of causality, in order to obtain a new strategy that can reveal new aspects about "activity" of the brain. To this purpose, we performed a comparison between the patterns of activation as detected by our GC-like measure and GLM.

The paper is organized as follows. Section 1.1 describes all the reasons why the detection of activations has to be seen a causality relation between the stimulus and the BOLD signal of each voxel. Section 2 contains material and methods used in the rest of the paper: the General Linear Model is used in a standard way to detect activation from fMRI data. The Granger Causality, currently used to study connectivity of the brain areas, tests a causality measure to detect activations; the Gini index, a measure of "sparsity" of data, is employed in connection to the localization of activations; a comparison between Granger Causality model and General Linear Model which demonstrate that GC is more general (as a model) than GLM. Section 3 describes the finger tapping experiment with which we test our approach on detecting activations, and gives necessary details about data analysis, including the algorithm used. In section 4 results of the experiment are displayed and in section 5 the discussion of these results is carried on. The major achievements from our point of view are:

- A unifying framework between the study of connectivity and activations, under the area of causality
- The reduction of signal-to-noise ratio with respect to the General Linear Model.
- A suggestion to derive a new definition of activation.

## 1.1 Activations as causality relations

In this section we want to show that the best way to look for a brain activation is to look for a causality relation between the time series of the BOLD signal and the response model of the stimuli. There has been no universally accepted definition of causality, so it would be futile to search for a unique causality measure. Most of the earlier research literature attempts to discuss unique causes in deterministic situations, and two conditions are important for deterministic causation:

**Necessity**: if X occurs, then Y must occur;

**Sufficiency**: if Y occurs, then X must have occurred.

However, deterministic formulation, albeit appealing and analytically tractable, is not in accordance with reality, as no real-life system is strictly deterministic (i.e. its outcomes cannot be predicted with complete certainty). So, it is more realistic if one modifies the earlier formulation in terms of likelihood (i.e. if X occurs, then the likelihood of Y occurring increases). Till 1970, the causal modeling was mostly used in social sciences. This was primarily due to a pioneering work by Selltiz et al. [Sel] who specified three conditions for the existence of causality:

- There must be a concomitant covariation between X and Y;

- There should be a temporal asymmetry or time ordering between the two observed sequences;

- The covariance between X and Y should not disappear when the effects of any confounding variables (i.e. those variables which are causally prior to both X and Y) are removed.

The first condition implies a correlation between a cause and its effect, though one should explicitly remember that a perfect correlation between two observed variables in no way implies a causal relationship. The second condition is intuitively based on the arrow of time. The third condition is problematic since it requires that one should rule out all other possible causal factors. Theoretically, there are potentially an infinite number of unobserved confounding variables available, yet the set of measured variables is finite, thus leading to indeterminacy in the causal modeling approach. In order to avoid this, some structure is imposed on the adopted modeling scheme which should help to define the considered model. The way in which the structure is imposed is crucial in defining as well as in quantifying causality.

The first definition of causality which could be quantified and measured computationally, yet very general, was given in 1956 by Wiener [Wie]: "For two simultaneously measured signals, if we can predict the first signal better by using the past information from the second one than by using the information without it, then we call the second signal causal to the first one."The introduction of the concept of causality into the experimental practice, namely into analyses of data observed in consecutive time instants, time series, is due to Clive W. J. Granger, the 2003 Nobel prize winner in economy. In his Nobel lecture [Gra1] he recalled the inspiration by the Wiener's work and identified two components of the statement about causality:

1. The cause occurs before the effect;

2. The cause contains information about the effect that is unique, and is in no other variable.

As Granger put it, a consequence of these statements is that the causal variable can help to forecast the effect variable after other data has been first used [Gra1]. This restricted sense of causality, referred to as Granger causality, characterizes the extent to which a process $X_t$ is leading another process $Y_t$, and builds upon the notion of incremental predictability. It is said that the process $X_t$

Granger causes another process $Y_t$ if future values of $Y_t$ can be better predicted using the past values of $X_t$ and $Y_t$ rather than only past values of $Y_t$.

Our belief is that brain activations resulting from presentation of stimuli must be defined as causal relations occurring between the time of presentation of stimuli and signal changes across different voxels, since:

- The claims 1. and 2. are satisfied by the pair stimulus/BOLD signal.

- Current methods to detect activations are indeed already applying causality principles.

Here we propose the use a causality measure to detect brain activation, since it is already used to analyze connectivity: in this fashion *the same* causality measure can be used to study activations and connectivity.

Activations of brain areas are not important in themselves: instead it is important how, and if, the activity (which in fMRI is detected by means of BOLD signal) is related to the stimulus. The relationship between the presentation of the stimulus and the BOLD signal clearly satisfy the two fundamental component of the statement given by Granger in his Nobel lecture [Gra1]:

1. ``The cause occurs before the effect'': undoubtedly the stimulus occurs before the effect that is produces in the BOLD signal. This must be true for physiological reasons, because every stimulus reaches first some of senses, and then is elaborated by the brain, from which the BOLD signal is registered.

2. ``The cause contains information about the effect that is unique, and is in no other variable'': the stimulus not only contains the information about the effect (the BOLD signal) that is unique, but indeed this information is just what we want to measure to assert that the voxel is activated as a consequence of the presentation of the stimulus.

## 2. Materials and methods

### *2.2 Generalized Linear Model*

The most used method to analyze fMRI data is the Generalized Linear Model (GLM). Here we provide a brief description of this standard method, in the light of a comparison with a causality measure.

A general linear model explains the response variables $Y_j$ in terms of linear combinations of explanatory variables $x_{j1}$, plus an error term:

$$\begin{pmatrix} Y_1 \\ \vdots \\ Y_j \\ \vdots \\ Y_J \end{pmatrix} = \begin{pmatrix} x_{11} & \cdots & x_{1l} & \cdots & x_{1L} \\ \vdots & \ddots & \vdots & \ddots & \vdots \\ x_{j1} & \cdots & x_{jl} & \cdots & x_{jL} \\ \vdots & \ddots & \vdots & \ddots & \vdots \\ x_{J1} & \cdots & x_{Jl} & \cdots & x_{JL} \end{pmatrix} \begin{pmatrix} \beta_1 \\ \vdots \\ \beta_l \\ \vdots \\ \beta_L \end{pmatrix} + \begin{pmatrix} \varepsilon_1 \\ \vdots \\ \varepsilon_j \\ \vdots \\ \varepsilon_J \end{pmatrix}.$$

The $\beta_i$ are the unknown parameters, and the error terms $\varepsilon_j$ are supposed to be i.i.d normal variables. In practical cases this system cannot be solved since the number of observations $J$ is higher than the number of explanatory parameters $L$, so some method to choose parameters that ``best fit" the data is required. Usually this fit is achieved by the method of *ordinary least squares*. If we denote with $\tilde{\beta}$ the fitted parameters, the ordinary least square method minimizes the residual sum-of-squares $\sum_{j=1}^{J} (Y_j - x_{j1}\tilde{\beta}_1 - \ldots - x_{jL}\tilde{\beta}_L)^2$. It is well known (see […]) that the least squares estimates are given by $\beta = (X^T X)^{-1} X^T Y$.

In an experiment with fMRI, the sequence of stimuli is coded with a binary string, in which the symbol *1* appears when the stimulus is present, and the symbol *0* otherwise. After that the theoretical BOLD response *r(t)* is generated by applying a finite impulse response function *h(t)* to the sequence of stimuli *S(t)*: $r(t) = \sum_{i=1}^{p} h(i) \cdot S(t-i)$ i.e. the stimulus timing is convolved with the impulse response function *h* (usually $h(t) = t^{8.6} e^{-t/0.547}$ [Coh]). Then the curve *Z(t)* is fitted to real data *Y(t)* minimizing the sum of square differences, adjusting the parameters $a, b, \beta \in \Re$:

$Z(t) = a + bt + \beta r(t)$     (4)

Here the parameters *a,b* represent a linear trend, while $\beta$ is the amount of the BOLD signal in real data: it tells us ``how much" BOLD signal can be computed through the convolution with the stimulus. Moreover, the baseline is computed as

$B(t) = a' + b't$,     (5)

adjusting the parameters $a', b' \in \Re$ to minimize the sum of square differences with respect to real data *Y(t)* again. Finally, if the model (4) fits data (statistically) better than the baseline model (5), the voxel is declared to be active.

*The General Linear Model as a causality measure.* The principle behind the definition of a causality relation can be stated as: the explanation of a (caused) variable is better explained taking into account also another (causing) variable, than considering only the caused variable? Now, what does ``to explain" means in this context, and how to measure the explainability, depends on the particular application we are studying. Notions of causality based on information theory often adopt the criterion by which ``explain better" means ``explain with less bits", understanding the fact that describing a variable with less bits is more efficient or convenient. In the Granger setting the measure of ``how much'' a variable causes another variable is placed in the error variance: if the variance of the error is significantly smaller in the model (6) than in (7), the variable *X* is causing the variable *Y*. The measure of the strength of the causality is then given as a function of error variances of models (6) and (7). There are many other ways of measuring the strength of the causality effect among variables, and of course, it doesn't exist the ``right'' one.

Considering now the General Linear Model, the parameter $\beta$ in the model (4) is exactly the measure, already used as such, of the amount of the signal explainable in terms of linear combinations of entries of the input timing into the observed BOLD signal. If the ``$\beta - value$'' is (statistically) high, then most of the BOLD signal is explainable in terms of the input timing (through the proposed linear model), and the voxel is declared to be activated by the input. In other words the parameter $\beta$ in (4) measures how much the knowledge of the input timing improves the knowledge of the BOLD signal. Here the causality principle precisely stands: the stimulus timing causes the BOLD signal if the latter can be better (i.e. with a sufficiently high $\beta$-value) explained with the stimulus timing, than using only a baseline model.

## 2.2 Granger Causality

The standard test of GC developed by Granger [Gra] to detect causality relations based on claims *1.* and *2.* is based on a linear regression model

$$Y_t = \sum_{k=1}^{L} b_{1k} Y_{t-k} + \sum_{k=1}^{L} b_{2k} X_{t-k} + \xi_t$$

where $\xi_t$ are uncorrelated random variables with zero mean and variance $\sigma^2$, *L* is the specified number of time lags, and *t =L+1, . . . , N*. The null hypothesis that $X_t$ does not Granger cause $Y_t$ is supported when $b_{2k} = 0$ for *k=1,...,L*, reducing Eq. (6) to

$$Y_t = \sum_{k=1}^{L} b_{1k} Y_{t-k} + \xi_t$$

As a measure of the causality between the time series $X_t$ and $Y_t$ we choose the standard quantity $f_{X \to Y} = 1 - \frac{R_2}{R_1}$, where $R_1$ is the residual sum of squares in Eq. (6) and $R_2$ is the residual sum of squares in Eq. (7). $f_{X \to Y}$ is smaller if $R_2$ is approximately equal to $R_1$, i.e. when the model (6), which takes into account the stimulus timing, fits the data approximately in the same way as the model (7), which does not take into account the stimulus timing.

This linear framework for measuring and testing causality has been widely applied not only in economy and finance (see [Gew] for a comprehensive survey of the literature), but also in diverse fields of natural sciences such as climatology or neurophysiology, where specific problems of multichannel electroencephalogram recordings were solved by generalizing the Granger causality concept to multivariate case [Bli,Kam].

## 2.3 Sparsity of data and Gini index

The concept of sparsity of data in this paper is used to determine whether the activations detected using the Granger-like causality are more "localized" than the responses detected with GLM. In our minds the greater sparsity (the technical synonymous of "localization") is associated with a more accurate activation detection in GC-like model, but this is due to a stronger link between the BOLD signal of a voxel and the stimulus. This aspect will be better analyzed in Results and Discussion.

There are many measures of sparsity. Intuitively, a sparse representation is one in which a small number of coefficients contain a large proportion of the energy. This interpretation leads to further possible alternative measures. Inequity of distribution is the same as sparsity. An equitable distribution is one with all coefficients having the same amount of energy, the least sparse distribution. Indeed, there are many measures of sparsity: we choose the Gini Index because it is the only one that satisfies all desirable characteristics of a sparsity measure given in the following:

1. If a quantity is added to a component and the same quantity is subtracted from another component, then sparsity decreases.
2. Sparsity is scale invariant
3. Adding a constant to each coefficient decreases sparsity.
4. Sparsity is invariant under cloning[1]
5. As one coefficient increases indefinitely, the distribution becomes as sparse as possible.
6. Adding component with zero coefficient to a distribution increases the sparseness.
7. The ordering of the coefficients is not important.
8. The sparsity of the coefficients is calculated using only the magnitudes of the coefficients

These criteria give rise immediately to the sparsest distribution being one with one component owning all the wealth and the least sparse being one with everyone having equal wealth.

In [Hur] it is shown that only one measure (among the one analyzed) satisfies all the properties above, namely the Gini Index. Given a vector $v = (v_1,...,v_N)$, which we are supposing with components ordered form the smallest to the largest (by property 7), the Gini index is given by

$$GI(v) = 1 - 2\sum_{k=1}^{N}\left(\frac{v_k}{\|v\|_1} \cdot \frac{N-k+1/2}{N}\right) \quad (8)$$

---

[1] If there is a twin population with identical distribution, the sparsity in one population is the same for the combination of the two

The Gini index was originally proposed in economics as a measure of inequality of wealth [Gin], and its usefulness as a measure of sparsity has been shown in [Ric,Ric1,Hur1,Hur2]

## *2.4 Granger causality truly generalizes GLM*

In GLM the $\beta-value$ in (4) provides a measure of the information stored in the stimulus timing useful to explain the BOLD signal. The comparison is made between the model (4), which comprehends the $\beta-value$, and the model (5), which models the baseline with a linear trend. Here we show that the regression model of Granger causality and the regression model of GLM are very similar (i.e. perform similar actions), and that the difference relies in the fact that Granger causality takes into account also the influence (i.e. the ``persistence'' or the ``inertia'') of the past states of the voxel in the explanation of the BOLD signal. From this point of view Granger causality is more general than GLM, since in GLM past states of voxel cannot influence future states.

To show our thesis, let's write more explicitly the equations that define the General Linear Model (i.e. eq. (4), substituting $r(t)$ with the expression (3), and eq. (5)), comparing them with equations defining the Granger Causality:

- General Linear Model

$$Z_t = a + bt + \sum_{k=1}^{p} \beta h_k S_{t-k} + \xi_t \qquad \text{(GLM alternative hypothesis)}$$

$$B_t = a' + b't + \xi'_t \qquad \text{(GLM null hypothesis)}$$

- Granger Causality

$$Z_t = \underbrace{a + bt + \sum_{k=1}^{p} \beta h_k S_{t-k} + \xi_t}_{GLM\ alternative\ hypothesis} + \sum_{k=1}^{L} c_k Z_{t-k} \qquad \text{(Granger Alternative Hypothesis)}$$

$$B_t = \underbrace{a' + b't + \xi'_t}_{GLM\ null\ hypothesis} + \sum_{k=1}^{L} c'_k Z_{t-k} \qquad \text{(Granger null hypothesis)}$$

Both in GLM and in the Granger causality test an *F* test can be used to distinguish if the alternative hypothesis is verified. Then the GLM takes the $\beta$ as a measure of the activation, while in the Granger setting the ratio of the estimated variances of errors $\frac{\text{var}(\xi_t)}{\text{var}(\xi'_t)}$ in the two models of alternative and null hypothesis.

But the main difference is that the Granger Causality setting includes also an autoregressive component both in the alternative and in the null hypothesis. This means that Granger Causality takes into account also the influence of some states of the voxel in detecting activation. In other words a voxel is declared to be activated if its BOLD signal is better explained taking into account the input sequence and the past BOLD signal itself, rather than the past BOLD signal only.

# 3. The experiment

*Subjects:* 12 right-handed healthy subjects (7M/5F, mean age± st.dev. = 21 ± 1 years) volunteered to participate. All subjects received medical, neurological and psychiatric examinations, routine laboratory tests and a structural MR scan of the brain to rule out any disorder that could affect brain function or metabolism. No subject had taken any psychotropic medication or any other medication for at least 4 weeks prior to the study. All subjects gave their written informed consent after the study procedures and potential risks had been explained (Protocol n. 020850 approved by the Ethical Committee of the University of Pisa).

*Image acquisition:* Functional magnetic resonance imaging (fMRI) images were acquired to explore brain activity elicited during a simple finger tapping motor task. We used a Gradient Echo echoplanar (GRE-EPI) sequence with a GE Signa 1.5 Tesla scanner (General Electric, Milwaukee, WI, USA) [repetition time (TR) 2000 ms, field of view (FOV): 24 cm, 25 axial slices, slice thickness 4 mm, echotime (TE): 40 ms, flip angle 90°, image plane resolution 64x64 pixels, voxel dimensions 3.75 x 3.75 x 4 mm]. For the finger tapping task, two sessions (time series), each consisting of 181 brain volumes, were obtained for each subject. High resolution T1-weighted spoiled gradient recall (SPGR) images were obtained for each subject to provide detailed brain anatomy during structural image acquisition.

*Experimental paradigm:* BOLD signal was measured while subjects performed a finger tapping motor task: after a 30 s resting phase, subjects were requested to alternatively perform a finger tapping for 12 s and rest for 30 s. Each run was made of 5 repetition of 30s-rest plus 12s-finger tapping, and each subject performed 2 runs.

## 3.1 Data Analysis

To optimize the comparison between the General Linear Model approach and the Granger-like Causality approach, data preprocessing and group analysis (one sample T-test) were identically performed with the AFNI package (http://afni.nimh.nih.gov). More in details the finger tapping data has been pre-processed by aligning each voxel time series to the same temporal and spatial origin (3dTshift and 3dvolreg AFNI programs), removing spikes (3dDespike AFNI program), and concatenating the time series of the two runs (3dTcat AFNI program). As group analysis, we performed a within group one sample T-test. The correction of the T-contrasts for multiple comparisons across whole brain was made using Monte-Carlo simulations run via AlphaSim in AFNI with a voxelwise threshold of 0.001.

**Description of the Granger-like Causality algorithm:** Roughly speaking, this algorithm implements the (directed) Granger causality, from the time series representing the stimulus timing to the time series of the BOLD signal of each voxel. A more detailed description follows:

1. The algorithm accepts as inputs: the stimulus timing time series $\{S_t\}$, the time series of the BOLD signal $\{Z_t\}$ of each voxel, the number of input and output delays that have to be considered in the model, and the TR;

2. - Parameters $a, b, b_k, c_k$ are estimated to minimize the sum of square differences between the time series $\{Z_t\}_{t=1}^{K}$

$$Z_t = a + bt + \sum_{k=1}^{p} b_k S_{t-k} + \xi_t + \sum_{k=1}^{L} c_k Z_{t-k}$$

and the time series of the BOLD signal of the voxel. Then the residual variance *ARXvar* is computed.

- Parameters $a', b', c'_k$ are estimated to minimize the sum of square differences between the time series $\{B_t\}_{t=1}^{K}$

$$Z_t = a' + b't + \sum_{k=1}^{L} c'_k B_{t-k}$$

and the time series of the BOLD signal of the voxel. Then the residual variance *ARvar* is computed.

- The number representing the ``strength'' of the causality is computed

$$f_{X2Y} = 1 - \frac{|ARX\,\text{var}|}{|AR\,\text{var}|}$$

3. The time series representing the stimulus and the time series of the voxel BOLD signal are bootstrapped *100* times, and the step *(3)* is applied to the bootstrapped time series. The residual variances are stored in the vector *n_distr_gr*.

4. A two-sided rank sum test of the null hypothesis that data in the vectors *fX2Y* and *n_distr_gr* are independent samples from identical continuous distributions with equal medians, against the alternative that they do not have equal medians. If the test rejects the null hypothesis *fX2Y* is assigned as the causality strength between the stimulus and the BOLD signal of the voxel. Otherwise this value is *0*.

**Description of the GLM algorithm:** The General Linear Model is implemented in its AFNI version, where an estimate of the $\beta$-coefficients, one for each voxel, is obtained through a statistical test for the difference between the model (4) and the model (5). For each voxel then the $\beta$-value is assigned for output.

# 4. Results

We performed a comparison between General Linear Model and our Granger-like Causality measure, in order to detect activation patterns resulting from finger tapping data set. An exemplification of the comparison is given in fig. 1.

The activations, as detected with General Linear Model and with our Granger-like Causality measure, show similar activation patterns: the two methods identified significant bilateral activations in premotor, primary motor and supplementary motor cortex, striatum and cerebellum, within the movement-related network [Gou]. While the patterns of activations as detected with the two methods significantly overlapped, the results obtained with a GC-based approach revealed a more localized pattern of activation (see also fig. 2). To give a quantitative measure of the localization of activation we adopt compute the Gini Index on the activation matrices. Since the preprocessing and the group analysis were the same, the Gini index provides a measure of the sparsity (that is a synonymous of localization here) of detected activation in both methods. We found that in the Granger-like Causality method the Gini coefficient of the activation matrix is $\approx 0.6$, and in the General Linear Model the Gini coefficient of the activation matrix is $\approx 0.7$. Recall that Gini index is equal to *0* represents total equity, while *1* represents maximal localization or sparseness.

**General Linear Model**

$$Z_t = a + bt + \sum \underbrace{\beta h_k}_{=b_k} S_{t-k} \quad \text{(GLM alt. hyp.)}$$

$$B_t = a' + b't \quad \text{(GLM null hyp.)}$$

**Granger-like Causality**

$$Z_t = \underbrace{a + bt + \sum b_k S_{t-k}}_{\text{GLM alt hyp}} + \sum c_k Z_{t-k} \quad \text{(Granger alt. hyp.)}$$

$$B_t = \underbrace{a' + b't}_{\text{GLM null hyp}} + \sum c'_k Z_{t-k} \quad \text{(Granger null hyp.)}$$

–

**Fig. 1** Here $S_t$ represents the stimulus timing, $Z_t$ and $B_t$ the time series to be fitted to the BOLD signal for respectively the alternative and the null hypothesis. The regression-based model of GLM is included in the GC-based model, as pointed out by underbraces. An autoregressive component (in blue in GC-based model) is absent in GLM.

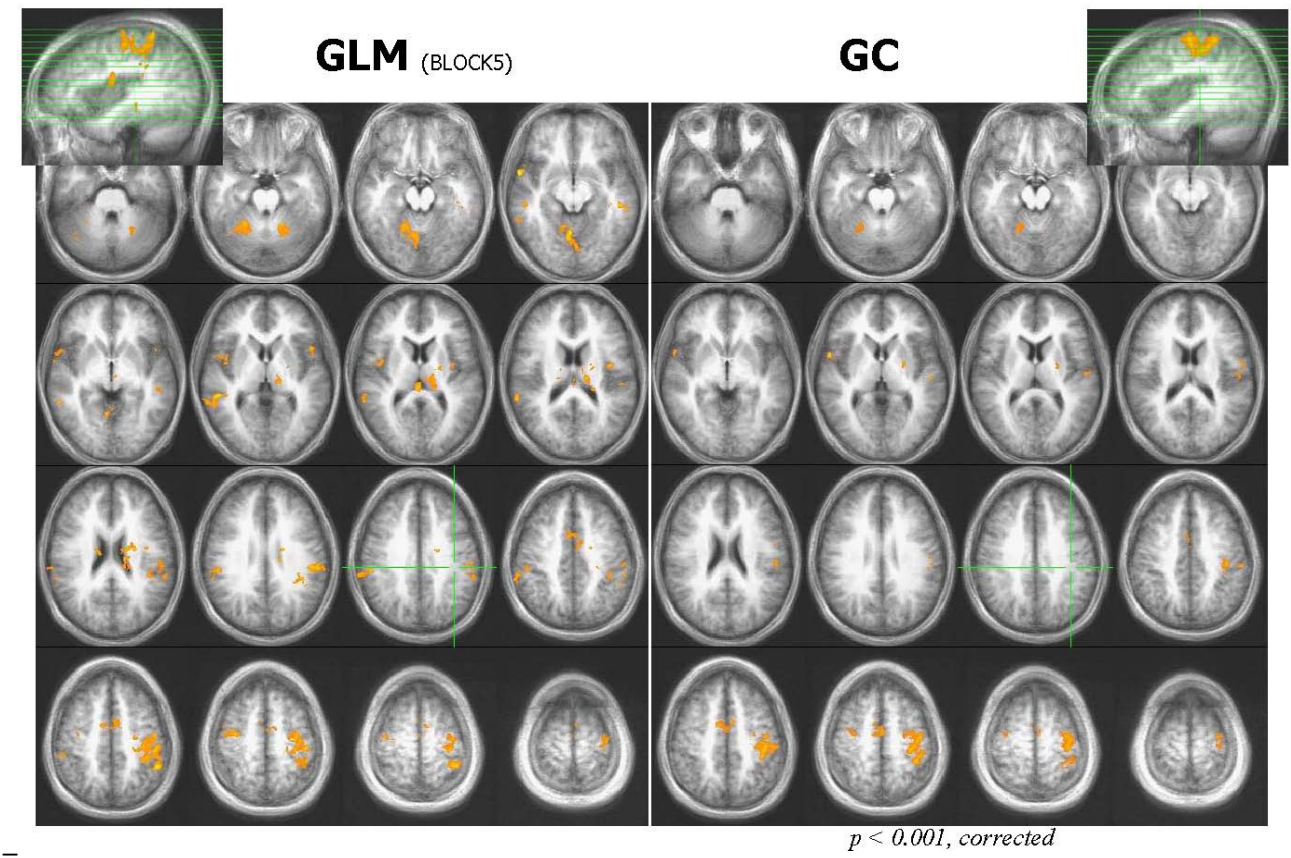

**Fig. 2** Activations patterns, as detected with GLM and GC-like measure, from the finger tapping data set. Both methods identified significant bilateral activations in premotor, primary motor and supplementary motor cortex, striatum and cerebellum, within the movement-related network.

# 5. Discussion

The main intent of this paper was to show that Causality measures serve both to detect activation and to analyze the connectivity. Here, we introduced a method to apply a Granger-like Causality measure to the detection of activations, since the use of causality measures – and in particular Granger Causality - in the context of connectivity analysis is well known (see for example […]). So the first immediate achievement, which is also the most innovative, is the possibility of *unifying activation and connectivity analysis*, within the framework of causality. Additionally, a better localization of the activation performed by the Granger/like Causality was obtained.

The first important point to be discussed is the *choice of the Granger Causality*. Our theoretical setting would allow to use any measure of causality, since any measure of causality could be a measure of the improvement of knowledge provided by one variable in knowing, or predicting, another variable. The particular choice of Granger causality is offered by a sizable number of reasons. The first one is that Granger causality (which is a mathematical object) is known and currently used in the analysis of fMRI signals, and indeed it is utilized also in many fields of economics and time series analysis. In this way the communication of the main finding (i.e. the ``unifying framework'' concept) is straightforward, without any critical technical issue in between. The second reason is the mathematical simplicity of Granger Causality, which is essentially a regressive model, and the fact that its model can be directly seen as a generalization of the General Linear Model (see section 3.3). The third reason, maybe the main one, is that Granger Causality is one of the primary tools to investigate connectivity between regions of the brain. This has as immediate consequence the ``unifying framework'' claim of the paper, since this work implies that Granger Causality can be used to analyze both activations and connectivity.

Another point to be discussed is the *choice of the data set*. It is better to specify at this point that the theoretical setting only would have been enough to assess all the claims of the paper. Indeed the equivalence between the concepts of activation of a voxel (as it is currently understood) and causality, and the fact that our Granger-like Causality generalizes the General Linear Method, is perfectly clear (and suffices) by the theoretical point of view. Thus our choice is to test the proposed method on a data set whose results are very simple, in such a way that a straightforward activation effect is immediately identifiable, and any other higher level effect is in fact unimportant. This allows a pure comparison between the detection of activations of the two techniques.

The most innovative achievement of our approach is the effect of *unifying the activation detection and the study of connectivity*, under the area of causality. These results support our theoretical model by proving a causal relationship between the stimulus and the BOLD signal of a voxel. Indeed we showed that Granger Causality can serve as well to detect activations in a finger tapping data set. The test confirmed that Granger Causality works well in detecting activations, and has also some advantages with respect to GLM (see below). Our work definitely shows that connectivity and activation analyses can indeed be treated as the same thing, the same problem: in detecting activations we are measuring causalities between the stimulus time series and each voxel, while in connectivity we are measuring causalities between the BOLD signal time series of a particular voxel and each other voxel. This has conceptual and practical consequences, since the activation and the connectivity analyses have been always seen as completely distinct approaches. The conceptual implication is of great significance, and deals with the concept of causality itself. A variable (a time series) causing another variable, in the modern acceptation, means that the knowledge of the first increases the knowledge (or the prediction goodness) of the other. But the literature on causality measures teaches that how vast can be the quantity of ``types'' of causality that can be of interest. It does not exist the ``right'' causality measure, and each causality measure reveals an aspect of the ``cause''. In the setting of activation detection it means that different causality measures will reveal different aspect of activations, and an immediate consequence is that the adjective ``active'' or ``non active'' are not enough: *a new concept of activity* should take into account different aspects of activations. On the other hand, a practical consequence of these considerations is immediate: the analyses of activation and connectivity are to be conducted in the same way, considering in the case of activations the presence of a sort of ``exogenous'' voxel whose time series is the stimulus timing.

As a supplementary result of our method, we showed that the Granger Causality, as a method to detect activation, is in fact more general that the General Linear Model. It is first demonstrated in

Section 3.3, where we put in evidence that the Granger Causality model ``includes'' the General Linear Model (see also fig. 1): the latter is obtained when the coefficients $c_k, c'_k$ are set to zero. In other words, the Granger Causality has an autoregressive component both in the alternative and in the null hypothesis that is not present in General Linear Model. So the meaningfulness in considering the stimulus sequence in the alternative hypothesis is reduced. We interpret this datum as a *reduction of the signal-to-noise ratio* in the Granger Causality method, because this takes into account also an autoregressive component. A practical way to observe this reduction is to compute the sparseness of activations obtained with the two methods: we associated the greater sparsity to a better localizations of activations, since we already clarified that the active brain areas are the same. The use of Gini index confirmed our theses.

Future works will address the analysis of activation patterns with different (and in particular non linear) causality measures, to reveal different aspects of activation of voxels, searching for a new concept of activation that can take into account all these aspects. The analysis of more complex data set, in order to exploit causality measures to evaluate the effect of higher order mental functions.